\documentclass[prl,aps,twocolumn,floatfix,showpacs]{revtex4}
\usepackage{graphics}
\begin{document}
\title{Two-band superfluidity from the BCS to the BEC limit}
\author{M. Iskin and C. A. R. S{\'a} de Melo}
\affiliation{School of Physics, Georgia Institute of Technology, Atlanta, Georgia 30332, USA}
\date{\today}

\begin{abstract}
We analyze the evolution of two-band superfluidity from the weak
coupling Bardeen-Cooper-Schrieffer (BCS) to the strong coupling
Bose-Einstein condensation (BEC) limit.
When the interband interaction is tuned from negative to positive values, 
a quantum phase transition occurs from a $0$-phase to a $\pi$-phase state, depending 
on the relative phase of two order parameters.
Furthermore, population imbalances between the two bands can be created
by tuning the intraband or interband interactions.
We also find two undamped low energy collective excitations
corresponding to in-phase and out-of-phase modes.
Lastly, we derive the coupled Ginzburg-Landau equations, 
and show that they reduce to coupled Gross-Pitaevskii equations for two
types of bosons in the BEC limit. 

\end{abstract}
\maketitle

A two-band theory of superconductivity was introduced by Suhl et al.~\cite{suhl}
in 1959 soon after the Bardeen-Cooper-Schrieffer (BCS) theory 
to allow for the possibility of multiple band crossings at the Fermi surface. 
This model has been applied to high-$T_c$ superconductors and $\rm {MgB_2}$,
where, in the latter case, experimental properties can be well described
by a two-band weak coupling BCS theory~\cite{mgb1,mgb2,mgb3,mgb4}. 
Unfortunately, interband or intraband interactions 
can not be tuned in these condensed matter systems,
and their properties can not be studied away from the BCS regime.
However, two-band fermions may also be produced experimentally with ultracold atomic
Fermi gases in optical lattices~\cite{kohl} or in single traps of several hyperfine states.
In this case, (intraband and interband) interactions may be tuned using Feshbach resonances
which allow for the study of the evolution of two-band superfluidity from the BCS to the
Bose-Einstein condensation (BEC) limit. The BCS to BEC evolution in the two-band problem is much richer than
the one-band case which has already been experimentally 
studied~\cite{kohl,coldatom1,coldatom2,coldatom3,coldatom4,coldatom5,coldatom6},
since additional interaction parameters (interband and intraband) may be controlled externally.

Furthermore, a two-band model may be used not only to describe experiments
involving several hyperfine states of the same fermion, but also may be
used to study two different fermionic species (e.g. $^6$Li and $^{40}$K).
In systems involving mixtures of two different alkali atoms,
simple one-band theories may not be sufficient to describe the interactions
between the two species of atoms, and two-band theories may be necessary to model
future experiments. Thus, due to recent developments and advances in atomic physics described
above, and in anticipation of future experiments, we describe here the BCS to BEC evolution
of two-band superfluids for all coupling strengths at zero and finite temperatures.

The main results of our paper are as follows. 
We show that a quantum phase transition occurs from a $0$-phase to a $\pi$-phase state
(depending on the relative phase of the order parameters of the two-bands) 
when the interband interaction $J$ is tuned from negative to positive values.
We found that population imbalances between the two bands can be created
by tuning intraband or interband interactions. In addition, we describe
the evolution of two undamped low energy collective excitations 
corresponding to in-phase phonon (or Goldstone) and out-of-phase exciton (finite frequency) modes. 
Near the critical temperature, we derive the coupled Ginzburg-Landau (GL) 
equations for a two-band superfluid, and show that they reduce 
to coupled Gross-Pitaevskii (GP) equations for two types of composite bosons 
in strong coupling. 

In order to obtain the results described above, we start from a
generalized Hamiltonian for multi-band superfluids with spin (pseudo-spin) singlet pairing 
\begin{eqnarray}
H &=& \sum_{n,\mathbf{k},\sigma}\xi_{n,\sigma}(\mathbf{k})a_{n,\sigma}^\dagger(\mathbf{k}) a_{n,\sigma}(\mathbf{k}) \nonumber \\
&-& 
\sum_{n,m,r,s,\mathbf{q}} V_{nm}^{rs} b_{nm}^\dagger(\mathbf{q}) b_{rs}(\mathbf{q}),
\label{eqn:hamiltonian}
\end{eqnarray}
where the indices $n,m,r$ and $s$ label different energy bands (or components), and
$\sigma$ labels spins (or pseudo-spins).
The operators
$
a_{n,\uparrow}^\dagger(\mathbf{k})
$
and 
$
b_{nm}^\dagger(\mathbf{q})= \sum_{\mathbf{k}} \Gamma_{nm}^*(\mathbf{k})
a_{n,\uparrow}^\dagger(\mathbf{k}+\mathbf{q}/2) 
a_{m,\downarrow}^\dagger(-\mathbf{k}+\mathbf{q}/2)
$
create a single and a pair of fermions, respectively.
The symmetry factor $\Gamma_{nm}(\mathbf{k})$ characterizes 
the chosen angular momentum channel, where
$
\Gamma_{nm}(\mathbf{k})=k_{nm,0}/(k_{nm,0}^2 + k^2)^{1/2}
$
is for the $s$-wave interaction in three dimensions.
Here, $k_{nm,0}\sim R_{nm,0}^{-1}$ sets the scale at small and large momenta,
where $R_{nm,0}$ plays the role of the interaction range.
In addition
$
\xi_{n,\sigma}(\mathbf{k}) = \epsilon_n(\mathbf{k}) - \mu_{n,\sigma},
$ 
where 
$
\epsilon_n(\mathbf{k})=\epsilon_{n,0} + k^2/(2M_n)
$
is the kinetic energy ($\hbar = 1$) and $M_n$ is the band mass of 
the fermions.

\begin{figure} [ht]
\centerline{\scalebox{0.45}{\includegraphics{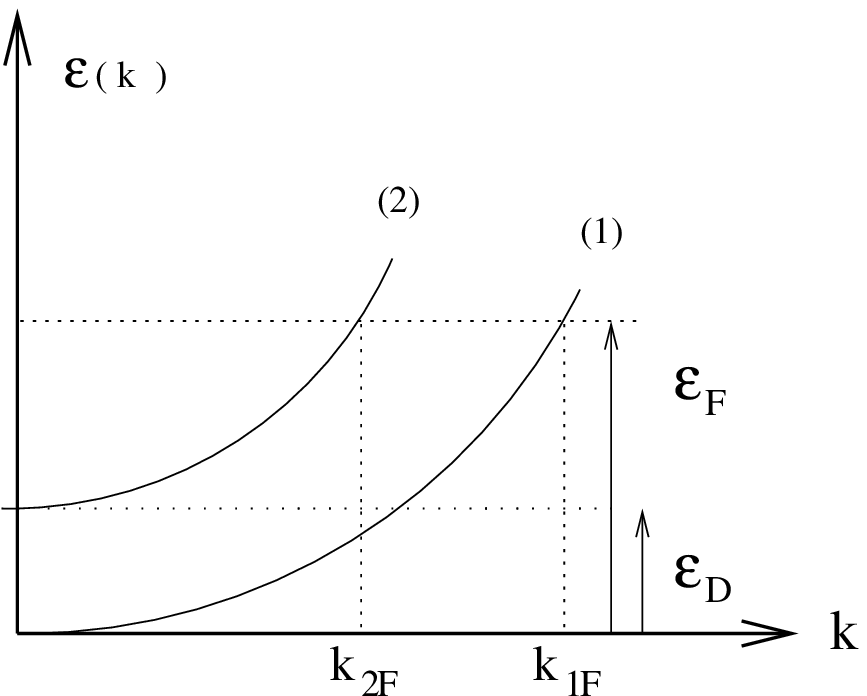}} 
\scalebox{0.45}{\includegraphics{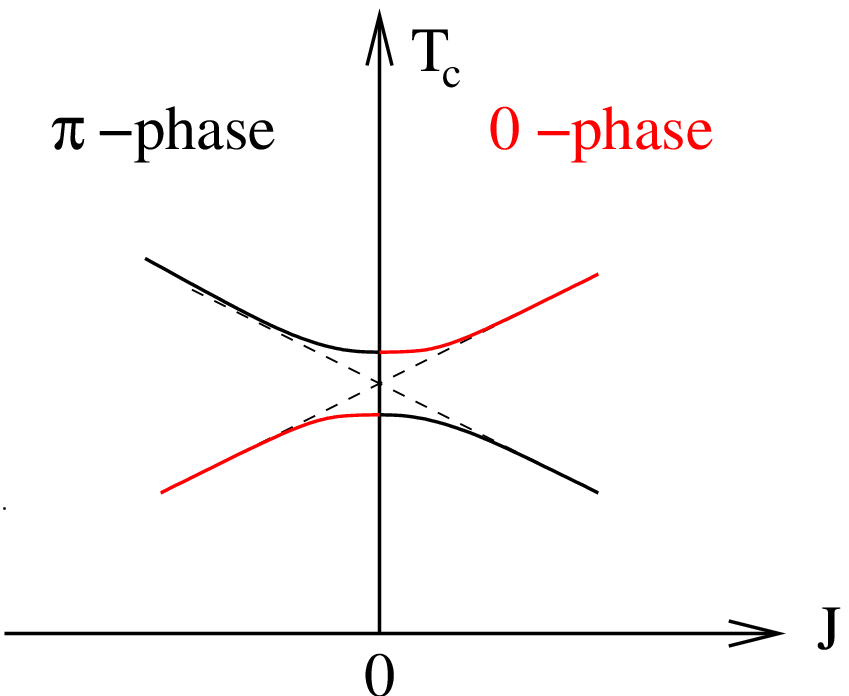}}}
\caption{\label{fig:bands} Schematic (a) figure of two bands with reference energies 
$\epsilon_{1,0} = 0$ and $\epsilon_{2,0} = \epsilon_D$, 
and (b) phase diagram of $0$-phase and $\pi$-phase states.
}
\end{figure}

From now on, we focus on a  two-band system
such that $V_{nm}^{rs} = V_{nr} \delta_{nm} \delta_{rs}$
with distinct intraband $(V_{11}, V_{22}) > 0$
and interband ($V_{12} = V_{21} = J$) interactions.
Notice that, $J$ plays the role of the Josephson 
interaction which couples the two energy bands.
In addition, we assume that the total number of fermions is fixed ($N = N_1 + N_2$) 
such that the chemical potentials of fermions are identical ($\mu_{n,\sigma} = \mu$), 
and that the reference energies are such that $\epsilon_{1,0} = 0$ and
$\epsilon_{2,0} = \epsilon_D > 0$, as shown in Fig.~\ref{fig:bands}a.
Here,
$
\epsilon_D = k_D^2/(2M_1) \le \epsilon_F = k_{1,F}^2/(2M_1),
$
where $\epsilon_F$ is the Fermi energy and $k_{n,F}$ are the Fermi momenta 
$k_{1,F} = k_F$ and $k_{2,F} = [2M_2(\epsilon_F - \epsilon_D)]^{1/2}$.
Since the low energy physics depends weakly on $k_{n,0}$ in dilute systems (${\cal N} R_0^3 \ll 1$)
characterized by $k_{n,0}^3 \gg k_{n,F}^3$ where ${\cal N} = N/{\cal V}$
is the density of fermions and ${\cal V}$ is the volume,
we assume for simplicity that $k_{n,0} = k_0 \gg k_F$.

The gaussian action for $H$ is ($k_B = 1, \beta = 1/T$)
\begin{equation}
S_{\rm gauss} = S_0 + \frac{\beta}{2} \sum_q \Lambda^\dagger (-q) \mathbf{F}^{-1}(q) \Lambda (q),
\label{eqn:action}
\end{equation}
where $q = (\mathbf{q}, iv_\ell)$ denotes both momentum and
bosonic Matsubara frequency $v_\ell = 2\ell\pi/\beta$.
Here, the vector
$
\Lambda^\dagger (-q) = \left[ \begin{array}{cccc} \Lambda_1^*(q), 
& \Lambda_1(-q), & \Lambda_2(-q), & \Lambda_2^*(q) \end{array} \right]
$
is the order parameter fluctuation field, and the matrix $\mathbf{F}^{-1}(q)$
is the inverse fluctuation propagator.
The saddle point action is
$
S_0 = -\beta \sum_{n,m} g_{nm}\Delta^*_{n,0}\Delta_{m,0}
+ \sum_{n, \mathbf{k}} \big \lbrace \beta 
[\xi_n(\mathbf{k}) - E_n(\mathbf{k})] 
- 2 \ln [1 + \exp(-\beta E_n(\mathbf{k})) ]
\big \rbrace,
$
where
$
g_{11} = - V_{22} / \det \mathbf{V},
$
$
g_{22} = - V_{11} / \det \mathbf{V}
$
and 
$
g_{12} = g_{21} = J / \det \mathbf{V}
$
with $\det \mathbf{V} = V_{11}V_{22} - J^2 > 0$.
Here,
$
E_n(\mathbf{k})=[\xi_n^2(\mathbf{k})
+ |\Delta_n(\mathbf{k})|^2]^{\frac{1}{2}}
$ 
is the energy of the quasiparticles and
$
\Delta_n(\mathbf{k}) = \Delta_{n,0}\Gamma_n(\mathbf{k})
$
is the order parameter.

The action given in Eq.~(\ref{eqn:action}) leads to the thermodynamic potential 
$
\Omega_{\rm gauss} = \Omega_0 + \Omega_{\rm fluct},
$
where $\Omega_0 = S_0/\beta$ is the saddle point and 
$
\Omega_{\rm fluct} = \beta^{-1} \sum_{q} \ln\det[\mathbf{F}^{-1}(q)/(2\beta)]
$
is the fluctuation contribution to $\Omega_{\rm gauss}$.
Expressing $\Delta_{n,0}$ in terms of its amplitude and phase
$
\Delta_{n,0} = |\Delta_{n,0}| \exp (i \varphi_n )
$
shows explicity the Josephson coupling energy
$
[ V_{22} |\Delta_{1,0}|^2 + V_{11} |\Delta_{2,0}|^2
- 2 J |\Delta_{1,0} \Delta_{2,0}| \cos (\varphi_2 - \varphi_1) ] / \det \mathbf{V}
$
of $\Omega_0$.
When $J > 0$, only the $0$-phase (or in phase)
$\varphi_2 = \varphi_1$ solution is stable. 
However, when $J < 0$, only the $\pi$-phase (or out of phase) 
$\varphi_2 = \varphi_1 + \pi$ solution is stable.
Thus, a phase transition occurs from the $0$-phase to
the $\pi$-phase when the sign of $J$ is tuned
from negative to positive values as shown in Fig.~{\ref{fig:bands}}b.

From the stationary condition $\partial S_{0}/\partial \Delta^*_n(q) = 0$,
we obtain the order parameter equation
\begin{eqnarray}
\left( \begin{array}{cc} O_{11} & O_{12} \\ O_{21} & O_{22} \end{array}\right)
\left( \begin{array}{c} \Delta_{1,0} \\ \Delta_{2,0} \end{array}\right) = 0,
\label{eqn:op}
\end{eqnarray}
where the matrix elements are given by
$
O_{nm} = - g_{nm} - \delta_{nm} \sum_\mathbf{k} |\Gamma_m(\mathbf{k})|^2 
\tanh[\beta E_m(\mathbf{k})/2] / [2E_m(\mathbf{k})].
$
Here, $\delta_{nm}$ is the Kronecker delta.
Notice that the order parameter amplitudes are the same for both the $0$-phase and 
$\pi$-phase as can be shown directly from Eq.~(\ref{eqn:op}),
but their relative phases are either $0$ or $\pi$. 
In what follows, we analyse only the $0$-phase state, keeping in mind that
analogous results (with appropriate relative phase changes) apply to the $\pi$-phase state.
We can elliminate $V_{nn}$ in favor of scattering length $a_{nn}$ via the relation
$
1/V_{nn} = - M_n {\cal V} / (4 \pi a_{nn}) + \sum_{\mathbf{k}} |\Gamma_n(\mathbf{k})|^2 /[2\epsilon_n(\mathbf{k})],
$
which can be solved to obtain
$
1/(k_{n,F} a_{nn}) = k_0/k_{n,F} - 4\pi/(k_{n,F} V_{nn} M_n {\cal V}).
$

The order parameter equation needs to be solved self-consistently with the
number equation $N = - \partial \Omega / \partial \mu$
leading to $N_{\rm gauss} = N_0 + N_{\rm fluct}$, and is given by
\begin{equation}
N_{\rm gauss} = \sum_{\mathbf{k}, \sigma, m} {\cal N}_{0,m} (\mathbf{k}) 
- \frac{1}{\beta} \sum_{q} \frac{\partial [\det \mathbf{F}^{-1}(q)] / \partial \mu}
{\det \mathbf{F}^{-1}(q)}.
\label{eqn:numbereqn}
\end{equation}
Here, the first term is the saddle point ($N_1 + N_2$) and the second term 
is the fluctuation ($N_{\rm fluct}$) contribution, where
$
{\cal N}_{0,m} = 1/2 - \xi_m(\mathbf{k}) \tanh[\beta E_m(\mathbf{k})/2] / [2E_m(\mathbf{k})] 
$
is the momentum distribution.
The inclusion of $N_{\rm fluct}$ is very important near the
critical temperature, however, $N_0$ may be sufficient at low temperatures~\cite{carlos,jan}.
Next, we discuss the $T = 0$ case.

In weak coupling ($\max\{ |\Delta_{1,0}|, |\Delta_{2,0}| \} \ll \epsilon_F$), 
the solutions of the order parameter equation are
$
\max\{|\Delta_{1,0}|, |\Delta_{2,0}| \} \sim 8\epsilon_F \exp[-2 + \pi k_0/(2k_F) - \phi_-]
$
and
$
\min\{|\Delta_{1,0}|, |\Delta_{2,0}| \} \sim 8\epsilon_F \exp[-2 + \pi k_0/(2k_F) - \phi_+],
$
while the number equation leads to $\mu \approx \epsilon_F$.
Here
$
\phi_\pm = \lambda_+ \pm [\lambda_+^2 - 1/\det \mathbf{\lambda} ]^{1/2}
$
where
$
\lambda_\pm = (\lambda_{11} \pm \lambda_{22})/(2\det \mathbf{\lambda}),
$
$
\det \mathbf{\lambda} = \lambda_{11}\lambda_{22} - \lambda_{12}\lambda_{21},
$
and $\lambda_{nm} = V_{nm} D_m$ are the dimensionless interaction parameters
with $D_m = M_m{\cal V}k_{m,F}/(2\pi^2)$ is the density of states per spin
at the Fermi energy.
On the other hand, in strong coupling ($\mu < 0$ and 
$\max\{|\Delta_{1,0}|, |\Delta_{2,0}|\} \ll |\mu| \ll \epsilon_0$),
the solution of the order parameter equations is
$
\mu = -\epsilon_0[\pi k_0/(2k_F \phi_+) - 1]^2,
$
while the number equation leads to
$
|\Delta_{m,0}|^2 = (8\pi {\cal N}_m/M_m)\sqrt{|\mu|/(2M_m)}.
$
Here, $\epsilon_0 = k_0^2/(2M_1)$ and ${\cal N}_m = N_m/{\cal V}$.
 Notice that the total density of fermions is
$
{\cal N} = {\cal N}_1 + {\cal N}_2 = (k_{1,F}^3 + k_{2,F}^3)/(3\pi^2).
$
The familiar one-band results are recovered when $J \to 0$
upon the use of the relation between $V_{nn}$ and $a_{nn}$. 
Next, we analyze the $T = 0$ evolution from BCS to BEC
for identical bands ($M_1 = M_2 = M$) with zero offset ($\epsilon_D = 0$).
For this purpose, we set $k_0 \approx 256 k_F$ and $V_{22} = 0.001$
in units of $5.78/(Mk_F{\cal V})$ [or $1/(k_F a_{22}) \approx -3.38$], 
and analyze two cases.

\begin{figure} [htb]
\centerline{\scalebox{0.35}{\includegraphics{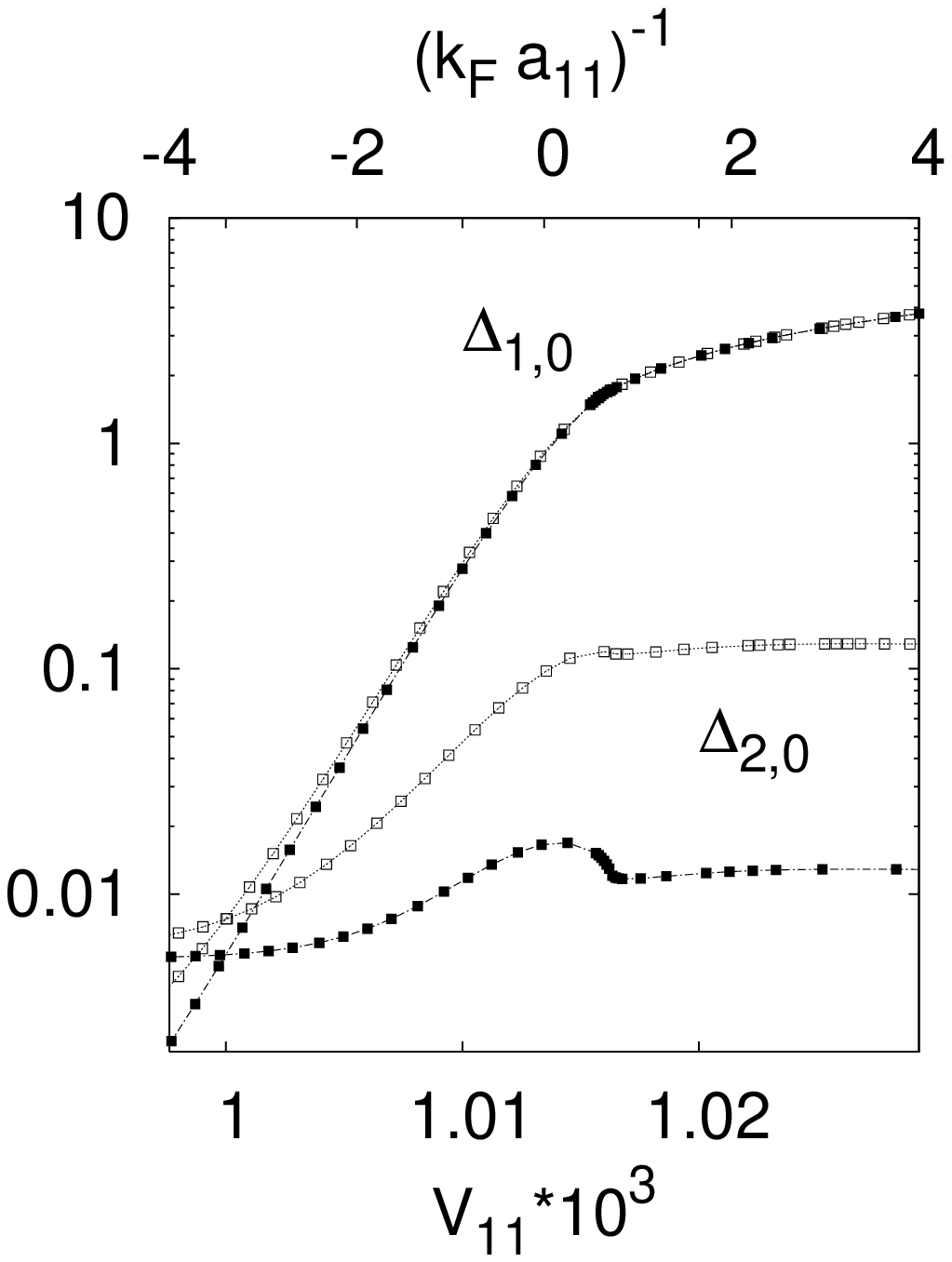} 
\includegraphics{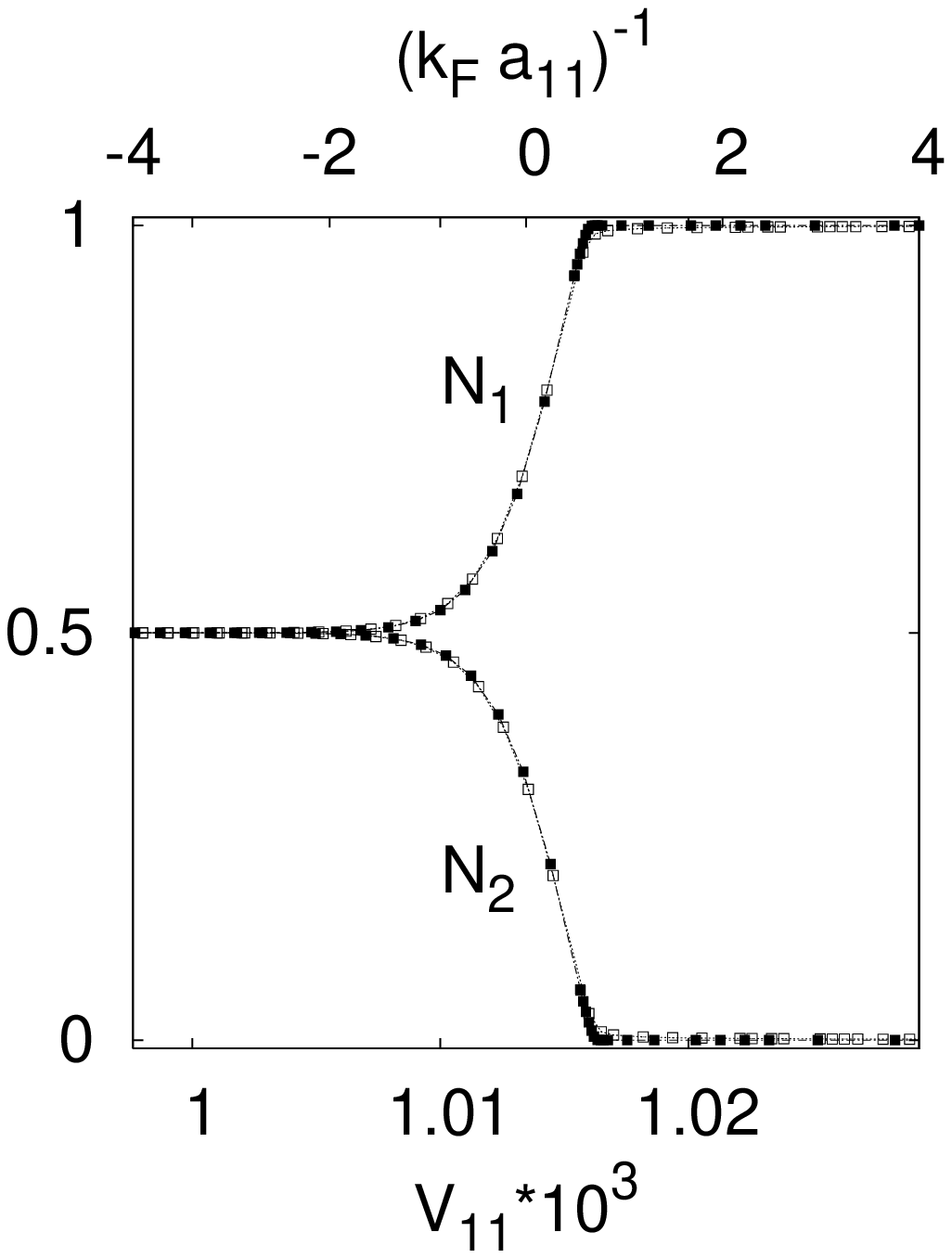}}}
\caption{\label{fig:gap.V11} Plots of
(a) order parameter amplitude $|\Delta_{n,0}|$ (in units of $\epsilon_F$), and
(b) fraction of fermions $N_n/N$ versus $V_{11}$ [in units of $5.78/(Mk_F{\cal V})$]
and versus $1/(k_Fa_{11})$ for $J = 0.001V_{22}$ (hollow squares) 
and $J = 0.0001V_{22}$ (solid squares).
}
\end{figure}

In the first case, we solve for $\mu$, $|\Delta_{1,0}|$ and $|\Delta_{2,0}|$ 
as a function of $V_{11}$ [or $1/(k_F a_{11})$], and show $\Delta_{n,0}$ 
in Fig.~\ref{fig:gap.V11}a for fixed values of $J$.
The unitarity limit is reached at $V_{11} \approx 1.0132V_{22}$ 
[or $1/(k_F |a_{11}|) \approx 0$] while
$\mu$ changes sign at $V_{11} \approx 1.0159$ [or $1/(k_Fa_{11}) \approx 0.69$].
The evolution of $|\Delta_{1,0}|$ is similar to the one band result~\cite{carlos} where it
grows monotonically with increasing $V_{11}$.
However, the evolution of $|\Delta_{2,0}|$ is non-monotonic where it has a hump
approximately $V_{11} \approx 1.0155V_{22}$ [or $1/(k_Fa_{11}) \approx 0.58$],
and it decreases for stronger interactions until it vanishes (not shown).
In Fig.~\ref{fig:gap.V11}b, we show that both bands have similar populations
for $V_{11} \sim V_{22}$. However, as $V_{11} / V_{22}$ increases,
fermions from the second band are transferred to the first, where
bound states are easily formed and reduce the free energy.

\begin{figure} [htb]
\centerline{\scalebox{0.35}{\includegraphics{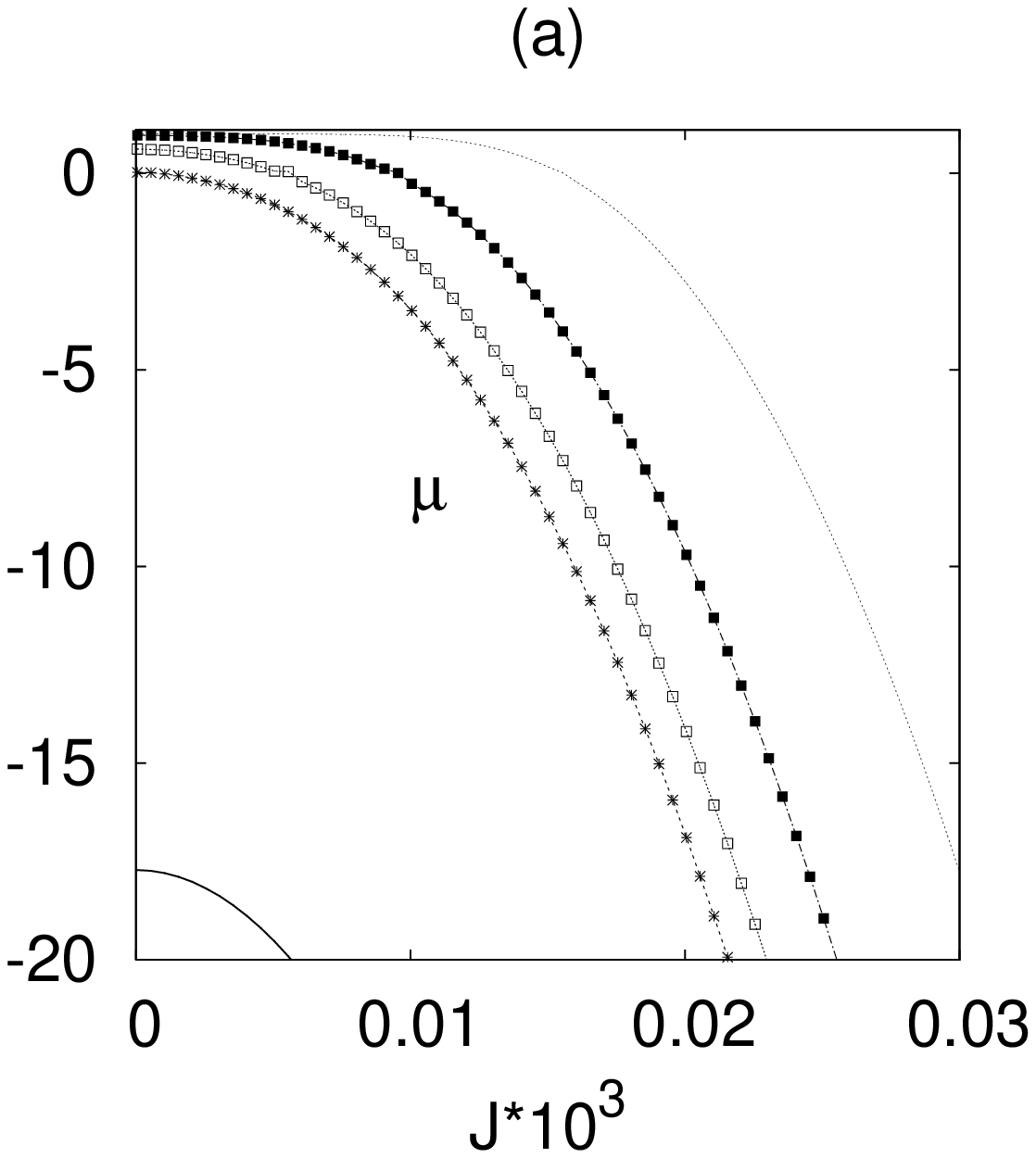} 
\includegraphics{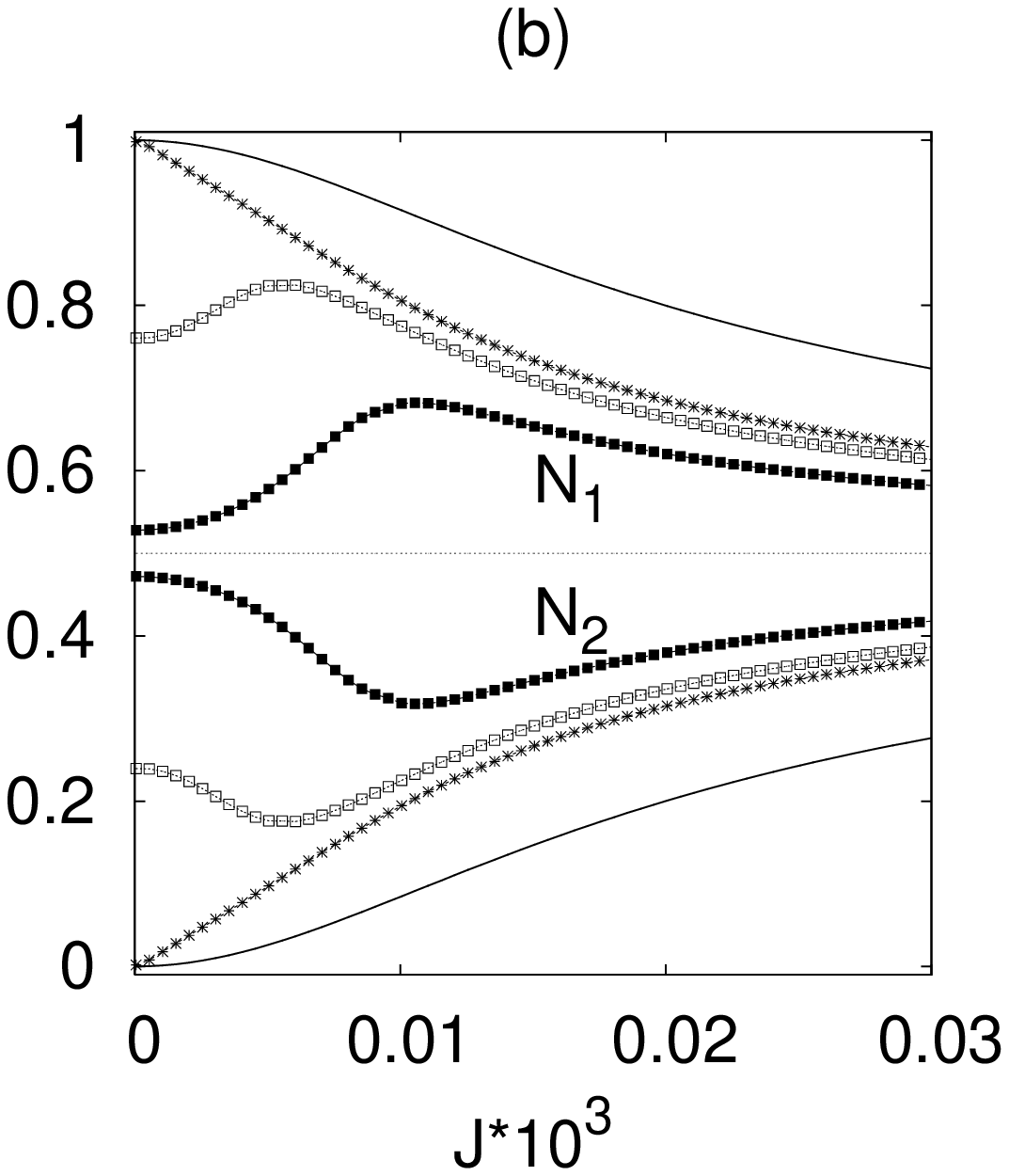}}}
\caption{\label{fig:gap.V12} Plots of
(a) chemical potential $\mu$ (in units of $\epsilon_F$), and
(b) fraction of fermions $N_n/N$
versus $J$ [in units of $5.78/(Mk_F{\cal V})$] for $V_{11} = \gamma V_{22}$ 
where 
$\gamma = 1$(dotted lines),
$1.010$ (solid squares),
$1.014$ (hollow squares),
$1.016$ (crossed lines) and
$1.030$ (solid lines);
or 
$1/(k_Fa_{11}) \approx -3.38, -0.81, 0.20, 0.70$ and $4.17$, respectively.
}
\end{figure}

In the second case, we solve for $\mu$, $|\Delta_{1,0}|$ and $|\Delta_{2,0}|$
as a function of $J$, and show $\mu$ in Fig.~\ref{fig:gap.V12}a for fixed values of $V_{11}/V_{22}$.
The order parameters $|\Delta_{1,0}|$ and $|\Delta_{2,0}|$ grow with increasing $J$ (not shown).
In Fig.~\ref{fig:gap.V12}b, we show that the band populations $N_1$ and $N_2$ for 
several values of $\gamma$. Notice the population imbalance and the presence of maxima (minima) in $N_1$ ($N_2$) for finite 
$J$, which is associated with the sign change of $\mu$ shown in Fig.~\ref{fig:gap.V12}a.
When $V_{11} > V_{22}$, there is population imbalance 
even for $J = 0$, because atom pairs can be easily transferred from the second band to the first until an optimal 
$J_o$ is reached. Further increase in $J$ produces also transfer from the 
first band to the second leading to similar populations for $J \gg J_o$.

Next, we discuss the low energy collective excitations at $T = 0$,
which are determined by $\det \mathbf{F}^{-1}(q) = 0$.
%
%
The phase-only collective excitations in weak and strong couplings lead to 
a Goldstone mode $w^2(\mathbf{q}) = v^2|\mathbf{q}|^2$ 
characterized by the speed of sound
$
v^2 = (D_1 v_1^2 + D_2 v_2^2) / (D_1 + D_2);
$
and a finite frequency mode $w^2(\mathbf{q}) = w_0^2 + u^2|\mathbf{q}|^2$
characterized by the finite frequency
$
w_0^2 = 4\alpha |g_{12}\Delta_{1,0}\Delta_{2,0}| \sqrt{\epsilon_F/|\mu|}
(D_1 + D_2) / (D_1 D_2)
$
and the speed
$
u^2 = (D_1 v_2^2 + D_2 v_1^2) / (D_1 + D_2).
$
In weak coupling, $\alpha = 1$ and
$
v_n = v_{n,F}/\sqrt 3,
$
while $\alpha = 1/\pi$ and
$
v_n = |\Delta_{n,0}|/\sqrt{8M_n|\mu|}
$
in strong coupling. Here, $v_{n,F}$ is the Fermi velocity.
It is also illustrative to analyse the eigenvectors associated with these solutions
in weak and strong couplings. 
In the limit of ${\mathbf q} \to 0$, we obtain
$
(\theta_1, \theta_2) \propto (|\Delta_{1,0}|, |\Delta_{2,0}|)
$
for the Goldstone mode corresponding to an in-phase solution, while
$
(\theta_1, \theta_2) \propto (D_2 |\Delta_{1,0}|, - D_1 |\Delta_{2,0}|),
$
for the finite frequency mode corresponding to an out-of-phase solution.
Our findings generalize Leggett's weak coupling results~\cite{leggett,iskin-twoband}.

Next, we discuss two band superfluidity near the critical temperature $T_c$,
where $|\Delta_{1,0}| \sim |\Delta_{2,0}| \to 0$.
For $T = T_c$, the order parameter equation reduces to
\begin{eqnarray}
\det \mathbf{O} = O_{11}O_{22} - O_{12}O_{21} = 0,
\label{eqn:tc}
\end{eqnarray}
and the saddle point number equation
$
N_0 = \sum_{\mathbf{k},n} n_F[\xi_n(\mathbf{k})]
$
corresponds to the number of unbound fermions, where
$n_F(x) = 1/[\exp(\beta x) + 1]$ is the Fermi distribution.
While $N_0$ is sufficient in weak coupling,
the inclusion of $N_{\rm fluct}$ is crucial in strong couplings,
and can be obtained as follows.
Near $T = T_c$, the fluctuation action $S_{\rm fluct}$ reduces to
$
S_{\rm fluct} = (\beta/2) \sum_{q,n,m} L_{nm}^{-1}(q) \Lambda_n^*(q) \Lambda_m(q)
$
where
\begin{equation}
L_{nn}^{-1} = - g_{nn} - \sum_{\mathbf{k}}
\frac{1 - n_F(\xi_{n+}) - n_F(\xi_{n-})} {\xi_{n+} + \xi_{n-} - iv_\ell} |\Gamma_n(\mathbf{k})|^2
\label{eqn:propagator}
\end{equation}
corresponds to the fluctuation propagator of band $n$,
$L_{n \ne m}^{-1} (q) = g_{nm}$, and $\xi_{n\pm} = \xi_n(\mathbf{k} \pm \mathbf{q}/2)$.
Thus, the resulting action leads to
$
\Omega_{\rm fluct} = \beta^{-1} \sum_{q} \ln[\det \mathbf{L}^{-1} (q)/\beta^2].
$
where $\det \mathbf{L}^{-1}(q) = L_{11}^{-1}(q) L_{22}^{-1}(q) - g_{12} g_{21}]$.
Notice that, $\det \mathbf{L}^{-1}(0) = 0$ also produces
Eq.~(\ref{eqn:tc}), which is the Thouless condition.
After the analytic continuation $iv_\ell \to w + i0^+$,
we expand $L_{nn}^{-1}(q)$ to first order in $w$ and second order in $\mathbf{q}$ such that
$
L_{nn}^{-1}(q) = a_n + \sum_{i,j}c_n^{ij} q_iq_j /(2M_n) - d_n w.
$
The time-independent coefficients are given by
$
a_n = - g_{nn} - \sum_{\mathbf{k}} X_n |\Gamma_n(\mathbf{k})|^2 / [2\xi_n(\mathbf{k})]
$
and
$
c_n^{ij} = \sum_{\mathbf{k}} \big\lbrace X_n\delta_{ij}/[8\xi_n^2(\mathbf{k})] 
- \beta Y_n\delta_{ij}/[16\xi_n(\mathbf{k})] 
+ \beta^2 X_n Y_n k_i k_j /[16M_n \xi_n(\mathbf{k})] \big\rbrace 
|\Gamma_n(\mathbf{k})|^2,
$
where $X_n = \tanh[\beta\xi_n(\mathbf{k})/2]$ and $Y_n = {\rm sech}^2[\beta\xi_n(\mathbf{k})/2]$.
Notice that, $c_n^{ij} = c_n\delta_{ij}$ is isotropic for the $s$-wave considered here.
The time-dependent coefficient has real and imaginary parts, and for the $s$-wave case is given by
$
d_n = \sum_{\mathbf{k}} X_n|\Gamma_n(\mathbf{k})|^2/[4 \xi_n^2(\mathbf{k})]
+ i(\beta \pi/8) \epsilon_0 D_n \sqrt{\mu/\epsilon_F} \Theta(\mu)/(\epsilon_0 + \mu),
$
where $\Theta (x)$ is the Heaviside function.
For completeness, we present the asymptotic forms of $a_n, c_n$ and $d_n$.
In weak coupling ($\mu \approx \epsilon_F$), we find
$a_n = - g_{nn} + D_n [\ln(T/T_c) + \phi_-]$,
$c_n = 7\epsilon_F D_n \zeta(3) / (12 T_c^2 \pi^2)$, and
$d_n = D_n [1/(4\epsilon_F) + i/(8T_c)]$,
where $\zeta(x)$ is the Zeta function, and $T_c$ is the physical critical temperature.
In strong coupling ($\epsilon_0 \gg |\mu| \gg T_c$), we find
$a_n = -g_{nn} - \pi D_n \epsilon_0/[2\sqrt{\epsilon_F} (\sqrt{|\mu|} + \sqrt{\epsilon_0})]$,
$c_n = \pi D_n / (16 \sqrt{|\mu|\epsilon_F})$ and
$d_n = \pi D_n / (8 \sqrt{|\mu|\epsilon_F})$.

In order to obtain $\Omega_{\rm fluct}$, there are two contributions, one from
the scattering states and the other from poles of $\mathbf{L}(q)$. The pole contribution 
dominates in strong coupling. In this case,
we evaluate $\det \mathbf{L}^{-1}(q) = 0$ and find the poles
$
w_\pm(\mathbf{q}) = A_+ + B_+|\mathbf{q}|^2 \pm  [(A_- + B_-|\mathbf{q}|^2)^2
+ g_{12}g_{21}/(d_1d_2)]^{1/2},
$
where
$
A_\pm = (a_1d_2 \pm a_2d_1)/(2d_1d_2) 
$
and
$
B_\pm = (M_2d_2c_1 \pm M_1d_1c_2)/(4M_1M_2d_1d_2).
$
Notice that, when $J \to 0$, we recover the limit of uncoupled bands with
$
w_n(\mathbf{q}) = a_n/d_n + |\mathbf{q}|^2 c_n/(2M_nd_n).
$ 
In the $\mathbf{q} \to 0$ limit, when $J > 0$ ($J < 0$),
the eigenvectors $[\Lambda_1^\dagger(0), \Lambda_2^\dagger(0)] = [g_{12}, a_1 - d_1 w_\pm (0)]$
correspond to an in-phase (out-of-phase) mode for $w_+(\mathbf{q})$ 
and an out-of-phase (in-phase) mode for $w_-(\mathbf{q})$.
Thus, we obtain $\Omega_{\rm fluct} = (1/\beta)\sum_{\pm, q} \ln[\beta(iv_\ell - w(\mathbf{q}))]$ 
which leads to
\begin{equation}
N_{\rm fluct} = \sum_{\pm, \mathbf{q}}
\frac{\partial w(\mathbf{q})}{\partial \mu} n_B[w(\mathbf{q})].
\label{eqn:nfluct}
\end{equation} 
For sufficiently strong couplings,
$\partial w_\pm(\mathbf{q}) / \partial \mu = 2$,
and the poles can also be written as 
$
w_\pm(\mathbf{q}) = - \mu_{B,\pm} + |\mathbf{q}|^2/(2M_{B,\pm}),
$
where $\mu_{B,\pm}$ is the chemical potential and $M_{B,\pm}$ is the
mass of the corresponding bosons.
In the case of identical bands with zero offset,
$c_1 = c_2 = c$ and $d_1 = d_2 = d$,
$\mu_{B,\pm} = - [a_1 + a_2 \pm \sqrt{(a_1 - a_2)^2 + 4g_{12}g_{21}}]/(2d)$ 
and $M_{B,\pm} = 2M$ in strong coupling.
Notice that the $+$ bosons always condense first  
for any $J$ (independent of its sign) since $\mu_{B,+} \to 0$ first.
Next, we analyze $T_c$ in weak and strong couplings.

In weak coupling, solutions to the order parameter equation Eq.~(\ref{eqn:tc}) are
$
T_{c,\mp} = (8\epsilon_F/\pi) \exp[\gamma -2 + \pi k_0/(2k_F) - \phi_\pm],
$
while the number equation Eq.~(\ref{eqn:numbereqn}) leads to $\mu \approx \epsilon_F$.
On the other hand, in strong coupling,
the solution of the order parameter equation is
$
\mu = -\epsilon_0[\pi k_0/(2k_F \phi_+) - 1]^2,
$
while the number equation ${\cal N}/2 = {\cal N}_{B,+} + {\cal N}_{B,-}$ with ${\cal N}_{B,+} \gg {\cal N}_{B,-}$ leads to
$
T_{c,+} = \pi \big \lbrace {\cal N}_{B,+}/[ \zeta(3/2) (M_{B,+}\sqrt{M_{B,+}}) ] \big \rbrace^{2/3},
$
since the $+$ bosons condense first.
Notice that, the physical critical temperature is
$T_c = \max \{T_{c,+}, T_{c,-}\}= T_{c,+}$ for any $J$.
Therefore, $T_c$ grows continuously from an exponential dependence on
interaction to a constant BEC temperature.

Next, we obtain the TDGL equations for $T \approx T_c$,
\begin{eqnarray}
\left[ a_n + b_n|\Lambda_n(x)|^2 \right.
&-& \left. \frac{c_n}{2 M_n} \nabla^2 - 
id_n\frac{\partial}{\partial t} \right]\Lambda_n(x) \nonumber \\
&+& g_{n \ne m} \Lambda_{m}(x) = 0,
\label{eqn:tdgl}
\end{eqnarray}
in the real space $x = (\mathbf{x},t)$ representation.
The coefficient of the nonlinear term
$
b_n = \sum_{\mathbf{k}} \big\lbrace X_n/[4\xi_n^3(\mathbf{k})] - \beta Y_n/[8\xi_n^2(\mathbf{k})] \big\rbrace 
|\Gamma_n(\mathbf{k})|^4
$
can also be obtained analytically in weak
$b_n = 7 D_n \zeta(3) / (8 T_c^2 \pi^2)$
and strong
$b_n = \pi D_n / (4 |\mu| \sqrt{2|\mu|\epsilon_F})$
coupling limits. 
In weak coupling, Eq.~(\ref{eqn:tdgl}) reduces to the coupled GL 
equations of two BCS type superconductors.
However, in strong coupling, it is more illustrative to derive TDGL equations
in the rotated basis of $+$ and $-$ bosons
$(\Phi_+^\dagger,\Phi_-^\dagger) = (\Lambda_1^\dagger,\Lambda_2^\dagger) \mathbf{R}^\dagger$, 
where $\mathbf{R}$ is the unitary matrix that diagonalizes the linear part of the TDGL equations.
In this basis, Eq.~(\ref{eqn:tdgl}) reduces to generalized GP equations 
of $\Phi_+$ and $\Phi_-$ bosons showing explicitly terms coming from density-density interactions 
such as $U_{\pm \pm} |\Phi_{\pm}|^2 \Phi_{\pm} $ or $U_{\pm \mp} |\Phi_{\pm}|^2 \Phi_{\mp}$.
In the case of identical bands with zero offset, this leads to 
$U_{++} = U_{--} = b/(2d^2)$ and $U_{+-} = U_{-+} = 3b/(2d^2)$
as the repulsive density-density interactions.

In conclusion, we showed that a quantum phase transition occurs from 
a $0$-phase to a $\pi$-phase state depending on the relative phase of the two order 
parameters, when the interband interaction $J$ is tuned from negative to positive values.
We found that population imbalances between the two bands can be created
by tuning intraband or interband interactions. In addition, we described
the evolution of two undamped low energy collective excitations 
corresponding to in-phase phonon (or Goldstone)
and out-of-phase exciton (finite frequency) modes.
Near the critical temperature, we derived the coupled Ginzburg-Landau (GL) 
equations for a two-band superfluid, and showed that they reduce 
to coupled Gross-Pitaevskii (GP) equations for two types of composite bosons 
in strong coupling. 

We thank NSF (DMR-0304380) for support.

\end{document}